\documentclass{elsart}
\usepackage{epsf}
\usepackage{rotating}

\catcode`@=11
\def\citer{\@ifnextchar[{\@tempswatrue\@citexr}{\@tempswafalse\@citexr[]}}

\def\@citexr[#1]#2{\if@filesw\immediate\write\@auxout{\string\citation{#2}}\fi
  \def\@citea{}\@cite{\@for\@citeb:=#2\do
    {\@citea\def\@citea{--\penalty\@m}\@ifundefined
       {b@\@citeb}{{\bf ?}\@warning
       {Citation `\@citeb' on page \thepage \space undefined}}%
\hbox{\csname b@\@citeb\endcsname}}}{#1}}
\catcode`@=12

\newcommand{\lsim}{\raisebox{-0.13cm}{~\shortstack{$<$ \\[-0.07cm] $\sim$}}~}
\newcommand{\gsim}{\raisebox{-0.13cm}{~\shortstack{$>$ \\[-0.07cm] $\sim$}}~}
\newcommand{\gau}{\tilde{\chi}}
\newcommand{\beq}{\begin{equation}}
\newcommand{\eeq}{\end{equation}}

\begin{document}

\hspace{0.9cm}\rightline{
        \begin{minipage}{4cm}
        PM/01--02\\
        hep-ph/0101177\hfill \\
        \end{minipage}        
}

\begin{frontmatter}
\title{Production of MSSM Higgs Bosons\\ in $\gamma\gamma$ Collisions}
\author{M.M.~M\"uhlleitner}
\address{Deutsches Elektronen-Synchrotron DESY, D--22603 Hamburg, FRG; \\
Laboratoire de Physique Math\'ematique et Th\'eorique, 
UMR5825--CNRS, Universit\'e de Montpellier II, F--34095 Montpellier Cedex 5,
France (present address)}
\begin{abstract}
  The heavy Higgs bosons $H,A$ of the minimal supersymmetric extension
  of the Standard Model can be produced as resonances in high-energy
  $\gamma\gamma$ colliders. Prospects of the search for these
  particles in $b\bar b$ and neutralino-pair final states are studied
  in this report. Heavy Higgs bosons can be found with masses up to
  about 70-80\% of the initial $e^+e^-$ collider energy for moderate
  values of $\tan\beta$, {\it i.e.} in areas of the parameter space
  not accessible at other colliders.
\end{abstract}
\end{frontmatter}

\section{Introduction}
%        ============
\vspace{-0.7cm}
The search for Higgs bosons is one of the most important endeavors of
present and future experiments. The minimal supersymmetric extension
of the Standard Model [MSSM] contains two isospin doublets of Higgs
fields which materialize, after electroweak symmetry breaking, in five
elementary Higgs particles \cite{habergun}: two neutral CP-even
($h,H$), one neutral CP-odd ($A$) and two charged ($H^\pm$) Higgs
bosons. The MSSM Higgs sector can be described, in leading order, by
two independent parameters which are in general chosen as the
pseudoscalar mass $M_A$ and $\tan\beta = v_2/v_1$, {\it i.e.} the
ratio of the two vacuum expectation values of the scalar Higgs fields.
While the mass of the lightest CP-even Higgs boson $h$ is bounded to
$M_h\lsim 130$~GeV \cite{mssmrad}, the masses of the heavy Higgs
bosons $H,A,H^\pm$ are expected to be of the order of the electroweak
scale up to about 1 TeV. The heavy Higgs bosons are nearly mass
degenerate. An important property of the SUSY couplings is the
enhancement of the bottom Yukawa couplings with increasing $\tan\beta$.
The negative direct search for the MSSM Higgs particles at LEP2 yields
lower limits $M_{h,H,A}\gsim 90$ GeV for the neutral Higgs masses
\cite{lep}. 
\vspace{-2mm}

Extensive studies have demonstrated that, while the light Higgs boson
$h$ of the MSSM can be found at the LHC, the heavy Higgs bosons $H,A$
may escape discovery for intermediate values of $\tan\beta$
\cite{lhc}. At $e^+e^-$ linear colliders heavy MSSM Higgs bosons can
only be found in associated production $e^+e^- \to HA$ \cite{3b}
according to the decoupling theorem. In the first phase of the TESLA
collider with a total $e^+e^-$ energy of 500 GeV the heavy Higgs
bosons can thus be discovered with masses up to about 250 GeV. To
extend the mass reach, the $\gamma\gamma$ option of TESLA can be used
in which high-energy photon beams are generated by Compton
backscattering of laser light \cite{plc}. Higgs particles can be
formed as resonances in $\gamma\gamma$ collisions \cite{borden}:
%\beq
$ \gamma\gamma \to h,H,A $.
%\eeq
%[For a recent list of references, see \cite{illa}.]  
Center-of-mass
energies of about 80\% of the $e^+e^-$ collider energy and a high
degree of longitudinal photon polarization can be reached at the
$\gamma\gamma$ collider. Integrated luminosities $\int {\mathcal L} =
300$~fb$^{-1}$ are expected {\it per annum} \cite{4a}.  Photon
colliders therefore provide a useful instrument for the search of
heavy Higgs bosons not accessible elsewhere \cite{4b}.

\vspace{-0.7cm}
\section{Analysis}
%        ========
\vspace{-0.7cm}
\underline{\it Branching ratios.}
%              =================
As promising examples for MSSM Higgs production in $\gamma\gamma$
collisions the two decay modes
%\beq 
$ H,A \to b\bar b$ and $\gau^0 \gau^0$
%\eeq
have been investigated for $\tan\beta=7$ and pseudoscalar masses 200
GeV $\leq M_A \leq 800$ GeV, {\it i.e.} the blind region at LHC. The
additional MSSM parameters for the gaugino sector have been chosen as
$M_2=-\mu=200$~GeV assuming a universal gaugino mass at the GUT scale.
They correspond to neutralino masses $m_{\gau^0_{1,2,3,4}} = 93, 161,
214, 256$~GeV and to chargino masses $m_{\gau^\pm_{1,2}} = 162,
258$~GeV.
\begin{figure}[hbt]
\vspace*{0.2cm}
\hspace*{0.0cm}
\epsfxsize=6.5cm \epsfbox{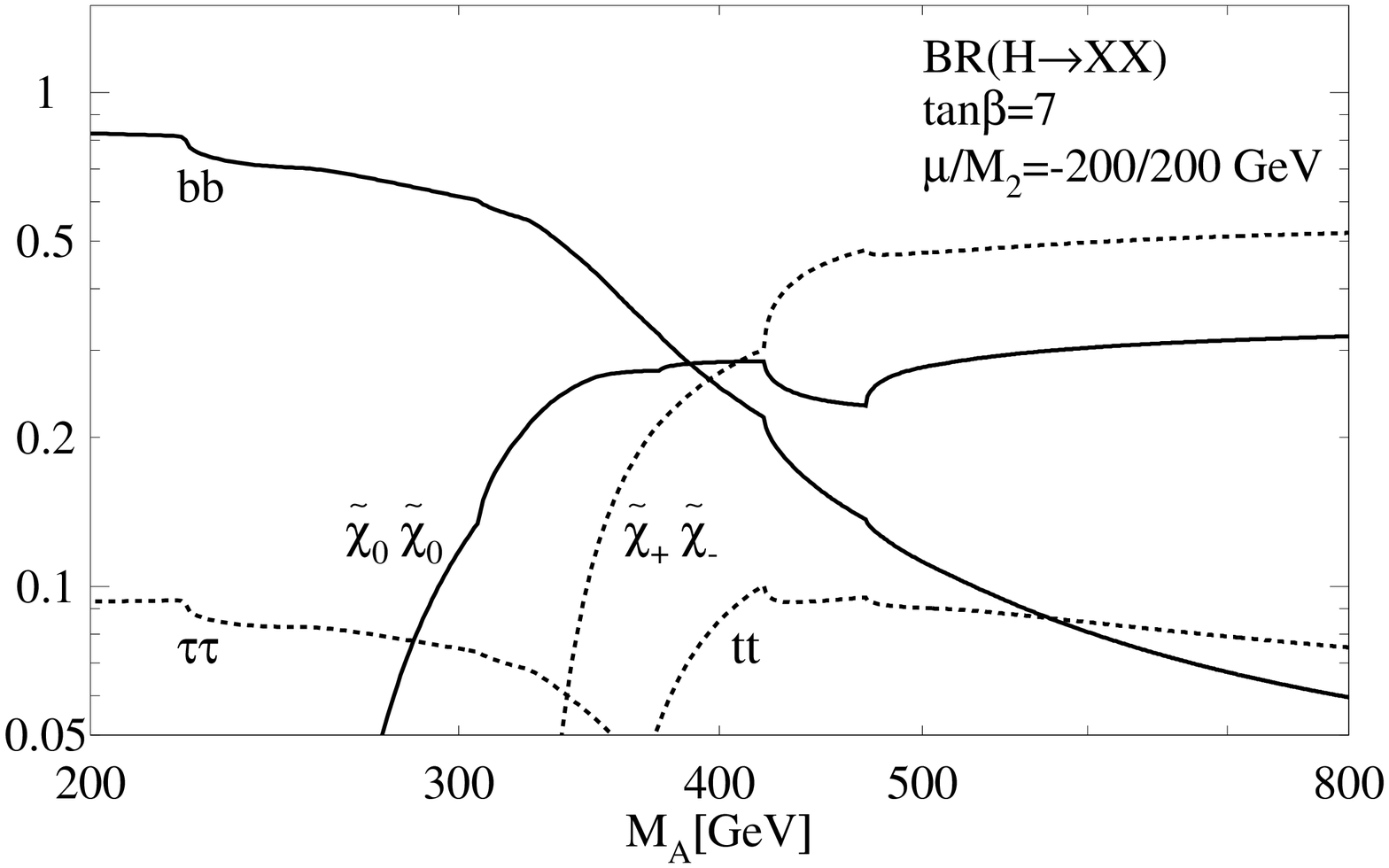}
\vspace*{-4.17cm}

\hspace*{7.2cm}
\epsfxsize=6.5cm \epsfbox{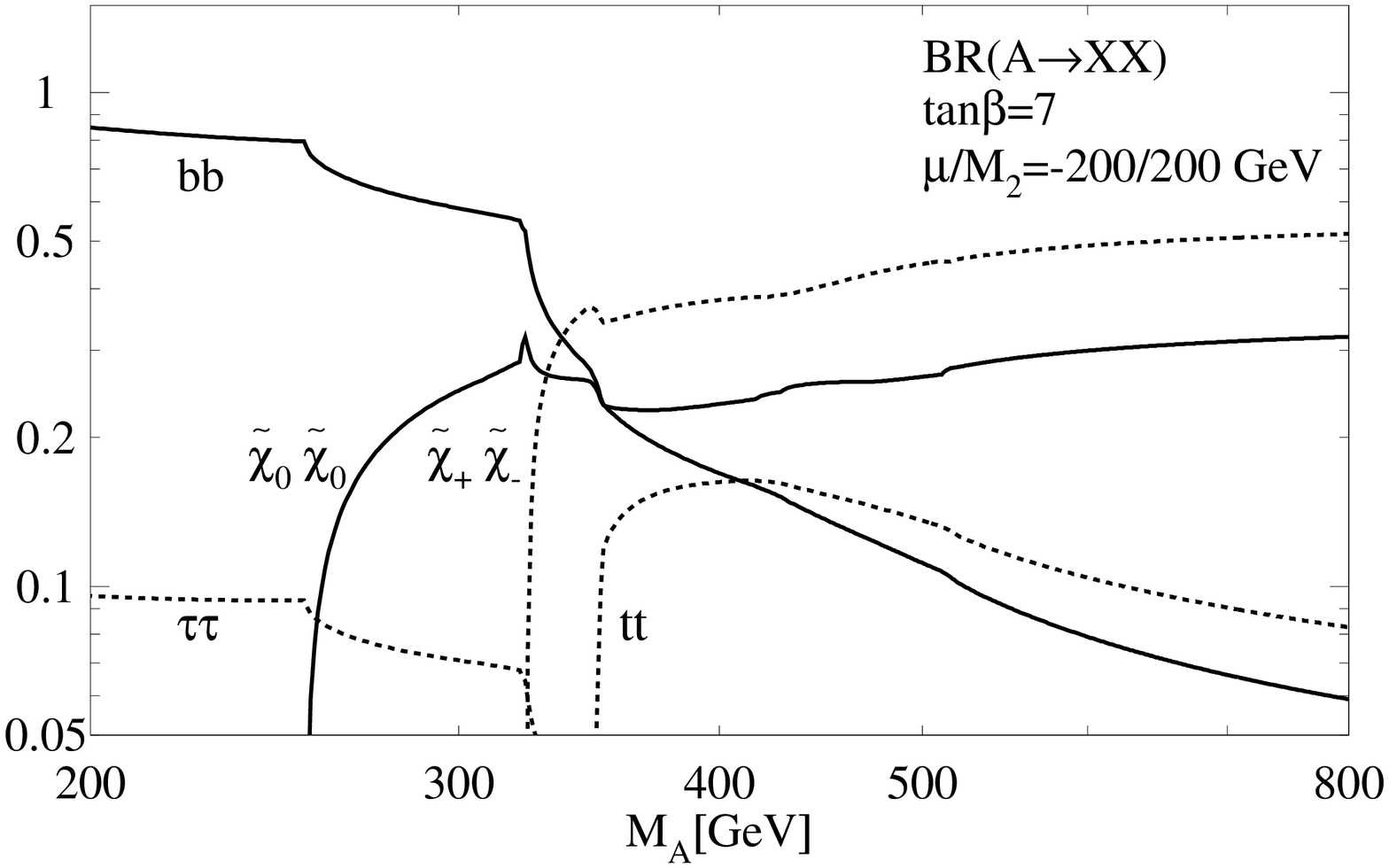}
\vspace*{-0.2cm}
\caption[]{\it \label{fg:br} Branching ratios of the heavy Higgs bosons
$H,A$ as a function of the corresponding Higgs mass. The MSSM parameters
have been chosen as $\tan\beta=7, M_2=-\mu=200$ GeV.}
\end{figure}
The branching ratios \citer{4c1,4c3} (excluding invisible decays into
$\gau^0_1\gau^0_1$ pairs) are presented in Fig.~\ref{fg:br}. For
moderate Higgs masses the $b\bar b$ decay modes turn out to be
dominant, while for large Higgs masses the dominant role is played by
the chargino and neutralino decay modes. The decays to $\tau^+\tau^-$
and $t\bar{t}$ pairs are suppressed with respect to $b\bar{b}$ decays 
by nearly an order of magnitude. 
\vspace{-2mm}

\underline{\it Signal.}
%              =======
At LO, the Higgs boson production in $\gamma\gamma$ collisions with
equal helicities is mediated by loop contributions of all charged
particles \cite{loop}. Sfermions, however, are assumed to be very heavy
and effectively decoupled in the present analysis. The NLO QCD
corrections consist of the two-loop corrections to the quark triangle
loops of the photonic Higgs couplings and the NLO corrections to the
final Higgs decays into quark-antiquark pairs ($b\bar b$). The
two-loop corrections to the Higgs couplings to photons are of moderate
size, if the quark mass inside the triangle loop is chosen as the
running quark mass at a typical scale fixed by the c.m.~energy of the
process \cite{hggqcd}. The NLO QCD corrections to the final Higgs
decays into bottom quarks generate large logarithms, which however can
be absorbed in the running Yukawa coupling at the scale of the
c.m.~energy \cite{4c2,hbbqcd}. After inserting the running quark masses,
the total QCD corrections increase the signal cross section by about
20--40\%.
\vspace{-2mm}

\underline{\it Backgrounds.}
%              ============
The main background processes $\gamma\gamma\to b\bar b$ and
$\gau^+\gau^-$ are pure QED reactions in lowest order. Since the
signals are generated for equal photon helicities, this configuration
can be enhanced by choosing opposite helicities for the incoming laser
photons and electrons/positrons in the initial state \cite{plc,kuehn}.
The NLO QCD corrections for polarized photon beams have been
calculated in Ref.~\cite{bkgqcd}.  For $b\bar b$ production they turn
out to be moderate for photons of opposite helicities but large for
photons of equal helicity. The differential cross section for
photons of equal helicity is suppressed by a factor $m_b^2/s$ at LO
due to a helicity flip of the bottom quark line. This suppression,
however, is removed by gluon radiation and the size of the cross
section increases to order $\alpha_s$.  Moreover, there are large
Sudakov and non-Sudakov logarithms due to soft gluon radiation and 
soft gluon and bottom exchange in the virtual corrections \cite{resum} 
which must be resummed. In order to suppress the gluon radiation we have
selected slim two-jet configurations in the final state, defined
within the Sterman--Weinberg criterion. If the radiated gluon energy
is larger than 10\% of the total $\gamma\gamma$ energy and if the
opening angle between all three partons in the final state is more
than 20$^o$, the event is classified as three-jet event and rejected.
The contamination of $b\bar{b}$ final states by the processes
$\gamma\gamma\to c\bar{c}$ can be kept under experimental control by
$b$ tagging.
\vspace{-2mm}

\underline{\it Interference.}
%              =============
The interference between the signal and background processes has been
taken into account properly. This part receives contributions only
from configurations with equal photon helicities in the initial state.
We have determined the NLO QCD corrections to quark final states
\cite{4b} including the resummation of the large
(non-)Sudakov logarithms. The QCD corrections to the interference term
are large.
\vspace{-0.7cm}

\section{Results and Conclusions}
%        =======================
\vspace{-0.7cm}
\underline{\it $b\bar b$ channel.}
%              ==================
We assume that a rough scan in the $e^+e^-$ energy will first be
performed which will provide preliminary evidence for the observation
of a Higgs boson in the $\gamma\gamma$ resonance channel. Since the
luminosity spectrum for photons of equal helicity is strongly peaked
at about 80\% of the $e^+e^-$ c.m.~energy \cite{plc,kuehn}, the
maximum of the spectrum will be tuned in the second step to the value
of the pseudoscalar Higgs mass $M_A$ at which the signal peak in the
energy scan appeared. The experimental analysis will be optimized
subsequently for the heavy Higgs boson search. A cut in the scattering
angle of the final bottom quark ($|\cos\theta | < 0.5$) strongly
reduces the background while it affects the signal process only
moderately. By collecting $b\bar b$ final states with a resolution in
the invariant mass $M_A \pm 3$ GeV, which is expected to be achieved
at the photon collider \cite{schreiber}, the sensitivity to the
combined $H$ and $A$ resonance peaks above the background is strongly
increased.

\begin{figure}[hbt]
\vspace*{0.2cm}
\hspace*{2.4cm}
\epsfxsize=9cm \epsfbox{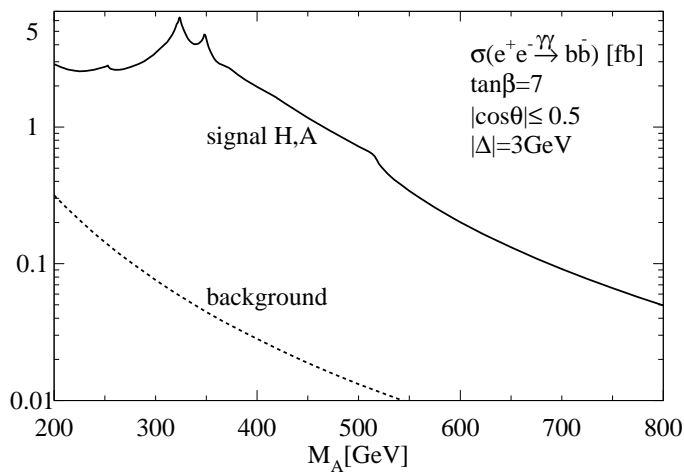}
\vspace*{-0.2cm}
\caption[]{\it \label{fg:bot} Cross section for resonant heavy Higgs boson
  $H,A$ production as a function of the pseudoscalar Higgs mass $M_A$
  with final decay into $b\bar b$ pairs and the corresponding
  background cross section. The maximum of the photon luminosity in
  the $J_z=0$ configuration has been tuned to coincide with $M_A$.
  The cross sections are defined in $b\bar b$ mass bins of $M_A\pm 3$
  GeV around the $A$ resonance. An angular cut on the bottom
  production angle $\theta$ has been imposed: $|\cos\theta|<0.5$. The
  MSSM parameters have been chosen as $\tan\beta=7, M_2=-\mu=200$
  GeV.}
\end{figure}
The result for the peak cross section is shown in Fig.~\ref{fg:bot} as
a function of the pseudoscalar mass $M_A$. It can be inferred from the
figure that the background is strongly suppressed against the signal.
Significances of the heavy Higgs boson signals sufficient for a
discovery of the Higgs particles up to about 70--80\% of the $e^+e^-$
c.m.~energy can be reached. At a 500 GeV $e^+e^-$ linear collider the
$H,A$ bosons with masses up to about 400 GeV can be discovered in the
$b\bar b$ channel at the photon collider, while for
$\sqrt{s_{ee}}$ above 800 GeV the range can be extended to about 600
GeV \cite{4b}. For heavier Higgs masses the signal rate
becomes too small for detection.
\vspace{-2mm}

\underline{\it $\gau^0\gau^0$ channels.}
%              =========================
For the MSSM parameters introduced above, the heavy Higgs bosons have
significant decay branching ratios to pairs of charginos and
neutralinos, cf.~Figs.~\ref{fg:br}. However, due to the integer
chargino charge, the chargino background from continuum production is
in general more than an order of magnitude larger than the signal.
Pairs of neutralinos cannot be produced in $\gamma\gamma$ collisions
at leading order so that neutralino decays open a potential discovery
channel for the heavy Higgs bosons $H,A$, see also Ref.~\cite{belan}.
This is obvious for moderate Higgs masses below the chargino decay
threshold. Detailed analyses of the topologies in the final state are
needed, however, to separate the neutralinos from the background
charginos above the threshold. For example, the signal channel
$\gamma\gamma\to H,A\to \gau_2^0 \gau_1^0$ leads, if the sfermions are heavy, 
in the $\gau^0_2$ cascade decay
predominantly to the hadronic final states $jj + E\!\!\!\slash$ 
whereas the continuum process $\gamma\gamma\to\gau^+_1\gau^-_1$ generates the
distinct final state $W^*W^*+E\!\!\!\slash$ with $jjjj+E\!\!\!\slash$ jet 
topologies. Thus, the neutralino decays
are expected to provide novel discovery channels of the heavy Higgs
bosons at a photon-photon collider.
\begin{figure}[hbt]
\vspace*{0.2cm}
\hspace*{2.4cm}
\epsfxsize=9cm \epsfbox{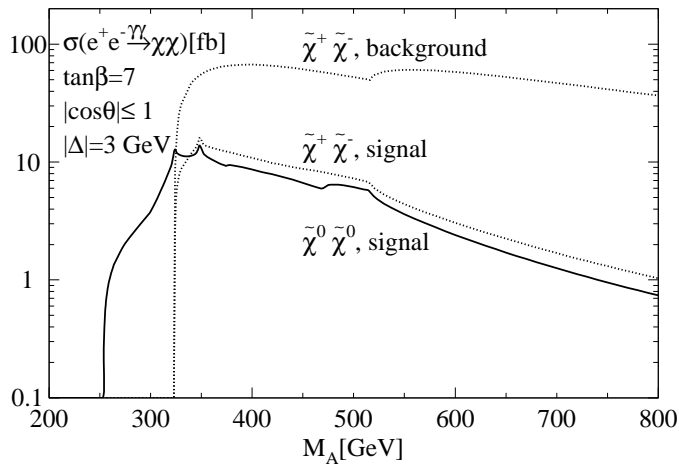}
\vspace*{-0.2cm}
\caption[]{\it \label{fg:gau} Same as in Fig.~\ref{fg:bot} but for
  chargino and neutralino final states.}
\end{figure}
\vspace{-2mm}

\underline{\it Summary.}
%              =======
It has been shown in this report that the heavy Higgs bosons $H$ and
$A$ of the minimal supersymmetric extension of the Standard Model MSSM
can be discovered for intermediate values of $\tan\beta$ up to masses of
about 400~GeV at a photon-photon collider in the first phase
of the linear collider project. The mass reach can be extended 
beyond 600~GeV at a TeV collider. The discovery potential in this region
of the supersymmetric parameter space is unique since neither at the
LHC nor in the respective $e^+e^-$ phase of the linear collider this region 
is accessible.

\noindent
{\bf Acknowledgements.}
Thanks go to M.~Kr\"amer, M.~Spira and P.M.~Zerwas for the fruitful
collaboration on the project. I am indebted to the organizers for the
invitation to the workshop.

\end{document}